\documentclass[12pt,a4paper]{article}
\usepackage{latexsym}
\usepackage{amsfonts}
\newcommand{\danger}[1]{\textbf{#1}}
\input{epsf}             
\usepackage[dvips]{graphicx}

\begin{document}
\title{\danger{Graphs on Surfaces and the Partition Function of String Theory}}

\author{\danger{J. Manuel Garc\'\i a-Islas\footnote{e-mail: jmgislas@gmail.com}}}

\maketitle

\danger{Abstract}. Graphs on surfaces is an active topic of pure
mathematics belonging to graph theory. It has also been applied to
physics and relates discrete and continuous mathematics. In this
paper we present a formal mathematical description of the relation
between graph theory and the mathematical physics of discrete string
theory. In this description we present problems of the combinatorial
world of real importance for graph theorists.

The mathematical details of the paper are as follows: There is a
combinatorial description of the partition function of bosonic
string theory. In this combinatorial description the string world
sheet is thought as simplicial and it is considered as a
combinatorial graph. It can also be said that we have embeddings of
graphs in closed surfaces.

The discrete partition function which results from this procedure
gives a sum over triangulations of closed surfaces. This is known as
the vacuum partition function.

The precise calculation of the partition function depends on
combinatorial calculations involving counting all non-isomorphic
triangulations and all spanning trees of a graph. The exact
computation of the partition function turns out to be very
complicated, however we show the exact expressions for its
computation for the case of any closed orientable surface.  We
present a clear computation for the sphere and the way it is done
for the torus, and for the non-orientable case of the projective
plane.

\section{\bf{Introduction}}
String theory is considered to be a quantum theory of all forces of
nature including of course quantum gravity \cite{gsw} \cite{p}.
When strings propagate over space-time they sweep a two dimensional surface known as world
sheet. The bosonic action of this wold sheet is proportional to its
area, and when propagating in Minkowski space-time it is given by
\begin{equation}
S_{P}= - \frac{1}{4 \pi \alpha^{'}} \int d\tau d\sigma \sqrt{-h}
h^{\alpha \beta} \partial_{\alpha}X^{\mu} \partial_{\beta} X^{\nu}
\eta_{\mu\nu}
\end{equation}
which is known as the Polyakov action. The partition function of
this theory for a fixed surface $\Sigma$ is given by

\begin{equation}
Z(\Sigma) = \int DX Dg \ e^{-S_{P}}
\end{equation}
where the integral is over all embeddings of the surface $\Sigma$ in
space-time(for example $R^n$), and over all metrics on the surface.

The combinatorial description of the string partition function is as
follows:

Consider the world sheet(two dimensional surface) $\Sigma$, and
triangulate $T$ it. Denote the vertices of the triangulated surface
by latin indices $i,j$.

Following \cite{adj}, the Polyakov action can be written in a
discrete form as

\begin{equation}
S(\Sigma_{T})(X) = \frac{1}{2} \sum_{i \sim j} (X_{i}-X_{j})^{2} +
\mu F(T)
\end{equation}
where $X_{i}, X_{j}$ denotes the image of vertices $i,j$ under the
embedding $X$ of the triangulated surface in space-time.

$i \sim j$ means the vertices $i$ and $j$ are joined by an edge,
$\mu$ is a parameter and $F(T)$ denotes the number of triangles of
the triangulation $T$. As explained in \cite{adj}, the analogous of
all metrics in the world-sheet corresponds in the discrete theory to
all non-isomorphic triangulations. Therefore the partition function
for a fixed topology $\Sigma$ is given by a sum over triangulations
\begin{equation}
Z(\Sigma) = \sum_{T} \int \prod_{i \in V(T)} dX_{i} \
e^{-S(\Sigma_{T})(X)}
\end{equation}
In \cite{adfo}, \cite{adf} this partition function was studied. In
this paper we now give a precise mathematical description which
stresses the mathematical side related to graph theory, and more
specifically to problems that graph theorist are interested in at
present and which are important for a deep understanding of the
partition function.

For example, in the above partition function sum it is evident that
we have to know how to generate all non-isomorphic triangulations of
an arbitrary two dimensional surface which is clearly understood in
graph theory. The procedure is to start with the irreducible
triangulations of a surface which number grows as the genus of the
surface grows. Then an arbitrary triangulation is always obtained
from the set of irreducible ones by certain moves known as vertex
splittings. The problem is that we do not know how many irreducible
triangulations there are for any surface. The most studied cases
have been until now, the sphere, the torus, the torus of genus 2,
the projective plane, the Klein bottle which are only a few cases.
It is certain that the problem becomes more difficult as the genus
of the surface grows. Recent studies on this direction have been
considered in \cite{az}. There has been also studies of similar
problem of finding embeddings of complete graphs on surfaces
\cite{abn}, \cite{grs}, \cite{kv}.

Let us mention a well known interesting combinatorial problem. When
we have a fixed triangulation of a world-sheet surface $\Sigma$, the
integral

\begin{equation}
Z(\Sigma_{T})= \int \prod_{i \in V(T)} dX_{i} \
e^{-S(\Sigma_{T})(X)}
\end{equation}
is related to the well known Matrix-Tree theorem of combinatorics
\cite{rl}.

When summing over different triangulations for a fixed closed
surface $\Sigma$, we will show how each summand of the partition
function is calculated and see that the number of spanning trees of
the triangulation is relevant. The matrix tree theorem tells us how
to calculate this number for any graph. The problem is that if the
graph has numerous vertices it is not very practical to use the
matrix tree theorem but an estimate number of the number of spanning
trees is needed for our calculations.

Besides for a fixed number of vertices there are many non-isomorphic
number of triangulations of a surface.

We divide this paper as follows. In section 2 we introduce the
discrete partition function of string theory and show how the
partition function is calculated for any closed surface. We will see
what each term of the sum is, even though this is not sufficient to
know what the sum converges to. More will be needed for that as we
will understand in this paper. In section 3 we introduce the
mathematical concept of graphs on surfaces and the way to generate
all triangulations of a surface from the irreducible set of
triangulations as well as the number of non-isomorphic
triangulations with a fixed number of vertices and the number of
spanning trees. In section 4 we go to the main part of the paper
which is to do explicit calculations on surfaces of the partition
function which was our motivation. We consider the cases of the
sphere which is the only one we can do more formally; we also show
how it is done(less formally but still with rigor) for the torus and
the projective plane; and finally in general the calculation for any
surface.

\section{The partition function}

In this section we describe the discrete partition function. The
nice thing about it is that it is completely combinatorial. Consider
first a closed vacuum string world-sheet embedded in a space-time of
dimension $D$. The sheet is a compact, connected two dimensional
surface without boundaries. Let $T$ be a non-degenerate
triangulation of it.\footnote{We give a formal mathematical
description of graphs on surfaces and in particular of
triangulations in section 4}. This means that $T$ itself can be seen
as a graph, i.e a finite collection of vertices and edges with the
following properties: for any two different vertices it can exist
one edge only which joins them; otherwise there is no edge between
two different vertices. Moreover, a single vertex can not be joint
to itself, i.e there are no loops. With these conditions we think of
the non-degenerate triangulation of the world-sheet surface as a
graph. Consider the discrete Polyakov action for a particular
surface $\Sigma$ and triangulation $T$ which is given by equation
(3).

Define the combinatorial Laplacian of a graph(which extends to our
triangulation) as follows

\begin{displaymath}
\Delta = \left\{ \begin{array}{ll}
d & \textrm{if $i=j$}\\
-1 & \textrm{if $i \sim j$}\\
0 & \textrm{otherwise}
\end{array} \right.
\end{displaymath}

where $d$ is the number of edges incident to a vertex which is known
as its valance. With this combinatorial Laplacian it is not
difficult to see that the discrete Polyakov action can be written as
\begin{equation}
S(\Sigma_{T})(X) = \frac{1}{2} \sum_{i \sim j} X_{i} \Delta  X_{j} +
\mu F(T)
\end{equation}
In this discrete theory as mentioned in the introduction, for a
fixed surface, all non-isomorphic triangulations play the role of
the metrics, and the maps which are defined on vertices, are just
the different embeddings of the triangulated sheet in space-time.
The partition function for a closed surface $\Sigma$ is given by

\begin{equation}
Z(\Sigma) = \sum_{T} \int \prod_{i \in V(T)} dX_{i} \
e^{-S(\Sigma_{T})(X)}
\end{equation}
where we sum over all non-isomorphic triangulations of the surface.
The most general partition function is given by summing over
different topologies
\begin{equation}
Z = \sum_{\Sigma} \sum_{T} \int \prod_{i \in V(T)} dX_{i} \
e^{-S(\Sigma_{T})(X)}
\end{equation}

Consider the integral (5) for a fixed triangulation $T$ of the
surface $\Sigma$. Let $v$ be any vertex of this triangulation $T$
and consider the graph $T-v$ which is given by considering the
complement of the vertex $v$ and of all the edges incident to it.

Let the image of vertex $v$, $X_{v}$ be fixed. The partition
function reduces then to an integral of over all embeddings of the
remaining vertices, with the requirement that the image $X_{v}$ is
fixed. Integral (5) up to a factor can be rewritten as
\begin{equation}
Z(\Sigma_{T}) = e^{- \mu F(T)} \int \prod_{i \in V(T)} dX_{i} \
e^{\frac{1}{2} \sum_{i \sim j} X_{i} \Delta_{(T-v)}  X_{j}}
\end{equation}
where $\Delta_{(T-v)}$ is the combinatorial Laplacian assigned to
the graph $T-v$ which can be obtained from the Laplacian $\Delta$
associated to the triangulation $T$ by removing from its associated
matrix the column and row labeled by the vertex $v$. The resulting
amplitude is up to a factor given by
\begin{equation}
Z(\Sigma_{T}) = e^{- \mu F(T)} \bigg( \frac{(2 \pi)^{N(V)-1}}{Det
\Delta_{(T-v)}} \bigg)^{\frac{D}{2}}
\end{equation}
where $N(V)-1$ is the number of vertices of the graph $T-v$, which
is just the number of vertices of the triangulation $T$ minus one
and $D$ is the dimension of the space-time in which our triangulated
surface lives.

The integral above in combinatorics is related to the Matrix-Tree
theorem \cite{rl} where it is described that  the determinant $Det
\Delta_{(T-v)}$ equals the number of spanning trees of the
triangulation $T$.\footnote{See appendix for a description of the
Matrix-Tree Theorem } The Laplacian can be generalized to a vertex
or edge weights description where the above integral is generalized,
however we do not describe it here.

Observe that for a non-degenerate triangulation $T$, the number of
spanning trees is clearly greater than one. Therefore as the
determinant appears as a denominator in the evaluation of the
partition function it is clear that the partition function
evaluation is bounded from above as
\begin{equation}
Z(\Sigma_{T}) < \bigg( (2 \pi)^{N(V)-1} \bigg)^{\frac{D}{2}}
\end{equation}
which comes from a degenerate case in which it could exist only one
tree.

The partition function is given by the sum over all triangulations

\begin{equation}
Z(\Sigma)= \bigg( {\frac{1}{2 \pi}} \bigg)^{\frac{D}{2}} \sum_{T}
e^{- \mu F(T)} \bigg( \frac{(2 \pi)^{N(V)}}{Det \Delta_{(T-v)}}
\bigg)^{\frac{D}{2}}
\end{equation}
The question now is how do we perform the above sum over all
triangulations for any arbitrary surface. This is what we do in
section 4 by considering the general orientable and non-orientable
case, giving details of the calculations for some examples. We also
need to know how are all triangulations of a surface generated. This
is what we describe in the following section.

\section{Graphs on surfaces: Triangulations}

We first give some definitions. A graph is a pair $G=(V(G),E(G))$
where $V(G) \neq \emptyset$ is called vertex set, and $E(G)$ is a
set where each element $e \in E(G)$ consist of a pair of elements of
$V(G)$. The elements of $E(G)$ are called edges. Two vertices are
said to be adjacent, if there is an element of $E(G)$ which joins
them. A graph with $n$ vertices is complete and denoted $K_n$ if any
two vertices are adjacent.

A graph $G$ is said to be embedded in a surface $\Sigma$ if the
vertices of $G$ are distinct points of $\Sigma$ and every edge of
$G$ is a curve in $\Sigma$ connecting the corresponding points.

A triangulation of a surface $\Sigma$ will be defined as an embedded
graph $T$ in the surface such that $\Sigma$ is divided into regions
called faces, such that each face is bounded by exactly three
vertices and three edges, and any two faces have either one common
vertex or one common edge or no common elements of the graph.

Two triangulations $T_1$ and $T_2$ are said to be isomorphic if
there is a one to one and onto mapping $\phi:V(T_1) \longrightarrow
V(T_2)$ such that $\phi(u)\phi(v) \in E(T_2)$ whenever $uv \in
E(T_1)$

Let $T$ be a triangulation of a surface $\Sigma$, and consider an
edge $e$ and their two triangles which contain it. Contract the edge
$e$ and replace the two double edges by single ones. This lead us to
a new triangulation see fig[1]. The inverse move is called vertex
splitting.

\begin{figure}[h]
\begin{center}
\includegraphics[width=0.3\textwidth]{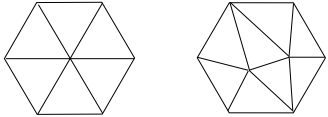}
\caption{Vertex splitting and edge contraction}
\end{center}
\end{figure}

Given a triangulation $T$ of a surface $\Sigma$ we can perform a
vertex splitting or an edge contraction in order to obtain a new
triangulation. When in a triangulation we cannot perform none edge
contractions which lead to a new triangulation again, we say that
our triangulation is minimal.

The sphere has only one 3-connected minimal triangulation given by
the embedded graph $K_4$ in the sphere \cite{mt}, \cite{s}.

And it is also known that all triangulations of the sphere are
obtained from the singular minimal triangulation $K_4$ \cite{s} by
vertex splittings.

Now it is known that there are two minimal triangulations of the
projective plane \cite{b} one given by the embedding of $K_6$ and
the other given in figure[2]. All the triangulations of the
projective plane are obtained from these two minimal triangulations
by vertex splittings.

\begin{figure}[h]
\begin{center}
\includegraphics[width=0.3\textwidth]{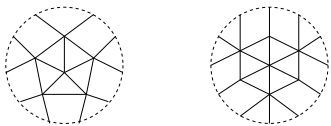}
\caption{Irreducible triangulations of the projective plane}
\end{center}
\end{figure}

Finally, for the torus it was shown \cite{l} that there are 21
minimal triangulations of it. For instance one is given by the
embedding of $K_7$ in the torus, $15$ triangulations with $8$
vertices, $4$ non-isomorphic ones with $9$ vertices and $1$
irreducible one with $10$ vertices, all of them non-isomorphic. And
from these 21 triangulations we can obtain all the triangulations of
the torus by vertex splittings moves.

It is known that the set of minimal triangulations for every surface
$\Sigma$ is finite \cite{be1}, \cite{be2} and the number grows
rapidly.

Given a graph $G$, a subgraph $H$ is given by $V(H) \subseteq V(G)$
and $E(F) \subseteq E(G)$. When it happens that $V(H)=V(G)$, $H$ is
called a spanning subgraph.

A tree is a connected graph without cycles. Given a graph $G$ we say
that $H$ is a spanning tree of $G$ if $H$ is a tree and a spanning
graph.

Given a graph $G$(or triangulation of a surface $T$), it is also
important to know the number of spanning trees of it, as will be
used in the next section. There is way to calculate the number of
different spanning trees of the graph by the matrix-tree theorem
given in the appendix.

However, if we need to know how the number of spanning trees grows
as the triangulation of a surface has more and more vertices(as is
needed for our calculations in the next section) the matrix-tree
theorem is not very useful for the purposes of computing the
partition function.

We therefore need to know a new way to calculate it which does not
require such a tedious calculation. Or we can try to give upper
bounds for this number. This is what we do in the following section,
we use an upper bound found in \cite{g}. As a mathematical problem
it will be interesting to have a better bound; or even better an
exact way to describe it.

\section{Computations of the partition function}

In this section we compute our partition function following all the
mathematical details we described in our previous section. Our
description is mathematically formal which gives a precision rule
for doing any calculation for any surface.

However it will be clear that even this combinatorial computations
are far from being trivial and when the genus of the surface grows
the computations become so difficult and completely unknown. This is
because we do not know the number of irreducible triangulations of
all closed surfaces, and some studies in this direction by finding
upper bounds have been studied in \cite{no} and recently approached
by \cite{az}.

Our next step is to show the way to perform this computation. The
thing is that we can only give an approximation of it, and give an
lower bound explicitly for the sphere only. Part of the calculation
can also be given for the case of the torus. The combinatorial
problem is complicated since as the number of spanning trees grows
when the triangulation has a larger number of vertices, the number
of non-isomorphic triangulations with a certain number of vertices
increases a lot as well. In fact this latter problem is a very
complicated one in the field of combinatorics. We proceed now to our
calculations. We denote a surface of genus $g$ by $S_{g}$.

\subsection{The Sphere}

In our notation we denote the sphere by $S_{0}$. Generally we saw
that the partition function was given by equation (12). We are
summing over triangulations, but it can be seen that such a sum can
be translated into a series sum over integers, as we now show.

As we have mentioned before, all of the triangulations of the sphere
can be obtained by refining a single simple triangulation which is a
minimal one \cite{s}. This minimal single triangulation of the
sphere is given by the complete graph $K_4$, that is, the
tetrahedron graph.

In the language of topological graph theory we say that the complete
graph $K_{4}$ is embeddable in the sphere. This graph is our first
summand of our partition function. From this single triangulation we
start taking vertex splittings.

It is clear that the following summands are given when we take
vertex splittings over and over; observe that the number of faces is
always even, that is, $N(F)=n=2k$, and the number of vertices is
giving by $N(V)=k+2$, which can be seen to be $k+\chi(S_0)$.

This lead us to rewrite the partition function sum (12) as follows

\begin{equation}
Z(S_{0}) = \bigg( {\frac{1}{2 \pi}}
\bigg)^{\frac{D}{2}}\sum_{k=2}^{\infty}  e^{- \mu 2k} C(T_{k+2})
\bigg( \frac{(2 \pi)^{k+2}}{\kappa(T_{k+2})} \bigg)^{\frac{D}{2}}
\end{equation}
where by $C(T_{k+2})$ we denote the number of non-isomorphic
triangulations with $N(V)=k+2$ vertices, and $\kappa(T_{k+2})$
denotes the number of spanning trees of a triangulation with
$N(V)=k+2$ vertices. For instance the first summand is given by only
one single graph which is $K_4$ where $C(T_4)=1$ and
$\kappa(T_{4})=16$.

Each summand has contributions from the number of non-isomorphic
triangulations with a fixed number of vertices and from the number
of trees of this triangulations.

The number of non-isomorphic triangulations of the sphere with a
fixed number of vertices has an asymptotic behavior \cite{t}. This
number is giving by

\begin{equation}
C(T_{k+2}) \sim \frac{1}{16} \bigg( \frac{3}{2 \pi}
\bigg)^{\frac{1}{2}} (k+2)^{- \frac{5}{2}} \bigg( \frac{256}{27}
\bigg)^{k+3}
\end{equation}
The number of spanning trees for two non-isomorphic triangulations
$T_1$ and $T_2$ with the same number of vertices($N(V)=k+2$), are
different since their Tutte polynomial invariant is different
\footnote{The number of spanning trees in a graph is a special case
of the Tutte Polynomial(see apendix)}. For this reason, the
calculation is harder than thought. The only thing we can do now is
to use an upper bound for the number of trees on any triangulation
with $N(V)=k+2$ vertices. As proved in \cite{g} any triangulation
with $k+2$ vertices has an upper bound for the number of spanning
trees given by

\begin{equation}
\kappa(T_{k+2}) \leq \frac{1}{k+2} \bigg( \frac{3(k+2)}{k+1}
\bigg)^{k+1}
\end{equation}
We therefore have

\begin{equation}
Z(S_{0}) \geq \bigg( {\frac{1}{2 \pi}}
\bigg)^{\frac{D}{2}}\frac{1}{16} \bigg( \frac{3}{2 \pi}
\bigg)^{\frac{1}{2}} \sum_{k=2}^{\infty} e^{- \mu 2k} (k+2)^{-
\frac{5}{2}} \bigg( \frac{256}{27} \bigg)^{k+3} \bigg(\frac{(2
\pi)^{k+2}(k+2)(k+1)^{k+1}}{(3(k+2))^{k+1}} \bigg)^{\frac{D}{2}}
\end{equation}
which tells us that we have a lower bound. It is now a task to study
its convergence for values of the parameter $\mu$ and of the
dimension $D$. It can be seen that the above partition function
converges for any value of $\mu \geq 2$.

Computations can be done with the help of a computer; Fix for
instance $\mu=2$, for $D=1$ we have $Z(S_0) \geq 0.5115676$; for
$D=2$ $Z(S_0) \geq 2.2794931$. It can be seen that for larger values
of $D$, the partition function goes to infinity, but also we have
that for larger values of $\mu$ the partition function converges
more rapidly. We therefore have that for a fixed value of the
dimension $D$, the partition function always converges for $\mu \geq
2$.

\subsection{The Torus and more surfaces}

As in the case of the sphere, all the triangulations of the torus
can be obtained by refining the minimal triangulations of it. In the
case of the sphere we had only one minimal triangulation. For the
case of the torus we have $21$ non-isomorphic minimal triangulations
\cite{l} from which we start in order to obtain all of the remaining
ones.

For instance the torus $S_{1}$ has as its simpler triangulation one
given by the embedding of the complete graph $K_{7}$. Therefore it
has $7$ vertices, $21$ edges and $14$ faces. We also have $15$
non-isomorphic triangulations with $8$ vertices, $4$ non-isomorphic
ones with $9$ vertices and $1$ irreducible one with $10$ vertices.

In all of these triangulations we can see that the number of faces
is always even, that is, $N(F)=n=2k$; we also have that $N(V)=k
=k+\chi(S_{1})$. This leads to the following sum

\begin{equation}
Z(S_{1}) = \bigg( {\frac{1}{2 \pi}}
\bigg)^{\frac{D}{2}}\sum_{k=7}^{\infty} e^{- \mu 2k} C(T_{k}) \bigg(
\frac{(2 \pi)^{k}}{\kappa(T_{k})} \bigg)^{\frac{D}{2}}
\end{equation}
where again $C(T_{k})$ denotes the number of non-isomorphic
triangulations with $N(V)=k$ vertices for the torus, and
$\kappa(T_{k})$ is the number of spanning trees of a triangulation
graph with $k$ vertices. The upper bound number of spanning trees is
the same we used before for the sphere since it is just a number
which depends on the number of vertices of the graph. But now our
problem is that the number of non-isomorphic triangulations
$C(T_{k})$ of the torus is not known in any way. It is just as
simple as noticing that we now have $C(T_7)=1, C(T_8)=15, C(T_9)=4,
C(T_{10})=1$. Then the sum above can be taken to the following
expression

\begin{eqnarray}
Z(S_{1}) = \bigg( {\frac{1}{2 \pi}} \bigg)^{\frac{D}{2}} \bigg[e^{-
14 \mu } \bigg( {\frac{(2 \pi)^7}{\kappa(T_{7})}}
\bigg)^{\frac{D}{2}} + 5 e^{- 16 \mu } \bigg( {\frac{(2
\pi)^8}{\kappa(T_{8})}} \bigg)^{\frac{D}{2}} + 20 e^{- 18 \mu }
\bigg(
{\frac{(2 \pi)^9}{\kappa(T_{9})}} \bigg)^{\frac{D}{2}} \nonumber \\
+ 21 \sum_{k=10}^{\infty} e^{- \mu 2k} C(T_{k}) \bigg( \frac{(2
\pi)^{k}}{\kappa(T_{k})} \bigg)^{\frac{D}{2}} \bigg]
\end{eqnarray}
where the major contribution is obviously given by

\begin{equation}
Z(S_{1}) \sim \bigg( {\frac{1}{2 \pi}} \bigg)^{\frac{D}{2}} 21
\sum_{k=10}^{\infty} e^{- \mu 2k} C(T_{k}) \bigg( \frac{(2
\pi)^{k}}{\kappa(T_{k})} \bigg)^{\frac{D}{2}}
\end{equation}
The real thing is that if we do not know anything about the number
$C(T_{k})$,  except for the irreducible triangulations, we cannot
compare the torus partition function to the sphere one.

However we would like to show only a partial comparison. This
partial comparison will be done by considering that there is only
one triangulation with a fixed number of vertices, for the sphere
and for the torus.

Suppose then that there is only one triangulation for the sphere
with $k+2$ vertices, that is $C(T_{k+2})=1$. Then

\begin{equation}
Z(S_{0})_{partial} \sim \bigg( {\frac{1}{2 \pi}}
\bigg)^{\frac{D}{2}}\sum_{k=2}^{\infty}  e^{- \mu 2k} \bigg(
\frac{(2 \pi)^{k+2}}{\kappa(T_{k+2})} \bigg)^{\frac{D}{2}}
\end{equation}
For the torus we have that there are 21 irreducible triangulations
from which we generate all triangulations. Suppose then that each of
the 21 irreducible triangulations generate only one respective class
of triangulations with a fixed number of vertices. We write

\begin{equation}
Z(S_{1})_{partial} \sim \bigg( {\frac{1}{2 \pi}}
\bigg)^{\frac{D}{2}} 21 \sum_{k=10}^{\infty} e^{- \mu 2k}  \bigg(
\frac{(2 \pi)^{k}}{\kappa(T_{k})} \bigg)^{\frac{D}{2}}
\end{equation}
both sums are partial but they still contain a sum over a very large
number of triangulations. The thing is that if we take $\mu \geq 2D$
we have that

\begin{equation}
Z(S_{0})_{partial} \gg Z(S_{1})_{partial}
\end{equation}
The above inequality is a very strict one and it tells us that the
partial contribution of the sphere is really much more bigger than
the partial contribution of the torus. Of course this is not telling
us that the original sums obey the same inequality, but the
interesting thing is the following. The number of non-isomorphic
triangulations with a fixed number of vertices for the torus, is
bigger than the one for the sphere with the same number of fixed
vertices. We have also mentioned that this number grows
exponentially when the genus of the surface grows. Therefore it is
expected that the inequality (22) changes when considering the
complete calculation.

It can also be suggested that partial contributions from other
topological surfaces are also dominated by the lowest genus surface.

Let us now give the partition function sum expression for any
orientable closed surface.

Observe first the following, which we assume happens for all of the
different topologies: The sums for the sphere and the partial sum of
the torus show that in the summands $2 \pi$ has exponent $k+
\chi(\Sigma)$, where $\chi(\Sigma)$ is the Euler characteristic of
the surface. We also have a factor which multiplies the summand
given by the number of non-isomorphic irreducible triangulations of
the surface, which for the sphere it was one and for the torus it
was $21$. The sum starts also from a higher number when the genus of
the surface increases. For the torus it starts for $k=10$ where $10$
is the number of vertices of the irreducible triangulation with more
vertices. The sum for any surfaces of any genus is given by

\begin{equation}
Z(S_{g}) = \bigg( {\frac{1}{2 \pi}} \bigg)^{\frac{D}{2}}
\sum_{k=n}^{\infty} e^{- \mu 2k} (C_{T_{k+\chi(S_{g})}})\bigg(
\frac{(2 \pi)^{k+ \chi(S_{g})}}{\kappa(T_{k+\chi(S_{g})})}
\bigg)^{\frac{D}{2}}
\end{equation}
where again we have that  $C(T_{k+  \chi(S_g)})$  denotes the number of non-isomorphic
triangulations with $N(V)=k+ \chi(S_g)$ vertices for the surface of genus $\chi(S_g)$, and
$\kappa(T_{k+\chi(S_g)})$  is the number of spanning trees of a triangulation
graph with $k + \chi(S_g)$ vertices.

\subsection{The Projective Plane and non-orientable surfaces}

With the calculation of the sphere and the way we explained how the
sum for the torus and any surface is to be obtained we could easily
know how the calculations follows for any non-orientable surface.
The only difference would be the appearance of the non-orientable
euler characteristic.

For instance recall that the projective plane has two irreducible
triangulations from which we can obtain all of its triangulations by
the vertex splitting moves. One is given by the embedding of the
complete graph $K_6$, with $6$ vertices, $15$ edges and $10$ faces.
Then each triangulation obtained from this irreducible one, by the
splitting moves, will have an even number of faces $2k$ and $k+1$
vertices where $k$ starts from $5$. The second irreducible
triangulation has $7$ vertices, $18$ edges and $12$ faces, and all
of the triangulations obtained from this irreducible one, will have
also an even number of faces $2k$ and $k+1$ vertices where $k$ stars
from $6$. Denote the projective plane by $N_{0}$. Therefore the
partition function is given by

\begin{eqnarray}
Z(N_{0}) = \bigg( {\frac{1}{2 \pi}} \bigg)^{\frac{D}{2}} \bigg[e^{-
10 \mu } \bigg( {\frac{(2 \pi)^6}{\kappa(T_{6})}}
\bigg)^{\frac{D}{2}}  + 2 \sum_{k=6}^{\infty} e^{- \mu 2k}
C(T_{k+1}) \bigg( \frac{(2 \pi)^{k+1}}{\kappa(T_{k+1})}
\bigg)^{\frac{D}{2}} \bigg]
\end{eqnarray}
We can easily guess and generalize the above sum to any
non-orientable surface of genus $g$. Denote such surface by $N_g$.
We therefore have the generalized partition function given by

\begin{equation}
Z(N_{g}) = \bigg( {\frac{1}{2 \pi}} \bigg)^{\frac{D}{2}}
\sum_{k=n}^{\infty} e^{- \mu 2k} (C_{T_{k+\chi(N_{g})}})\bigg(
\frac{(2 \pi)^{k+ \chi(N_{g})}}{\kappa(T_{k+\chi(N_{g})})}
\bigg)^{\frac{D}{2}}
\end{equation}
where as for the orientable case
 $C(T_{k+  \chi(N_g)})$  denotes the number of non-isomorphic
triangulations with $N(V)=k+ \chi(N_g)$ vertices for the non-orientable surface of genus $\chi(S_g)$, and $\kappa(T_{k+\chi(N_g)})$  is the number of spanning trees of a triangulation
graph with $k + \chi(N_g)$ vertices.

\section{Conclusion}

We have seen in this paper that there is a need to understand deeper
a pure mathematics problem in order to have a complete calculation
of the partition function of any two dimensional surface. In order
to have a complete sum over all triangulations of a surface we
learnt that we need to know first all the non-isomorphic irreducible
triangulations of the surface.

The problem clearly would be to have an asymptotic expression for
the number of non-isomorphic triangulations of any surface. Until
now, we have this expression for the sphere only \cite{t}. And it is
even very hard to find at least the number of irreducible
triangulations of a surface. There have been only upper bounds for
the number of irreducible triangulations of a surface of genus
$\chi(S_g)$ \cite{az}, \cite{no}. Even finding non-isomorphic
complete graph orientable or non-orientable  embeddings of complete
graphs on surfaces gives a huge number of families \cite{abn},
\cite{grs}, \cite{kv}.

Therefore the problem of computing partition functions for any
surface is incomplete. We therefore have that the discrete
formulation which we presented here, is not an advantage over the
continuous evaluations. It will be an advantage if we first solve
the combinatorial problems we presented here.

\bigskip

\bigskip

\danger{ACKNOWLEDGMENTS}

I want to thank Isidoro Gitler for very useful conversations related to combinatorics
and for pointing me to references.

\bigskip

\appendix
\section{The spanning trees of a triangulation}
This appendix describes the matrix-tree theorem. This is in order to
just understand how it is used in the paper. For a deeper
description of it see \cite{bb}.

 Let
$G$ denote a connected graph with vertex set $V(G)$ and edges set
$E(G)$. The combinatorial Laplacian $\Delta_{G}$ for the graph $G$
is defined in section 2, and it is given by a square matrix indexed
by their vertices. This square matrix is completely symmetric and
has determinant zero. Given any vertex $v$ of $G$ consider the
cofactor $\Delta_{G-v}$ of the matrix Laplacian $\Delta_{G}$ given
by deleting from $\Delta_{G}$ the row and column indexed by the
vertex $v$.

\bigskip

\danger{Matrix-Tree Theorem}. The determinant $Det(\Delta_{G-v})$ is
independent of the vertex $v$ and equals the number of spanning
trees of $G$.

\bigskip

There is also a generalization of the matrix-tree theorem when
considering graphs with edge weights. The number of spanning trees
of a graph can be thought as an invariant of the graph. This is
because this number is a particular case of a more general invariant
associated to graphs via a polynomial discovered by Tutte \cite{t2}.

The Tutte polynomial of a graph is a two variable one $T(G;x,y)$
which is defined by the contraction-deletion rule.

1.- If $G$ has no edges then $T(G;x,y)=1$

2.- $T(G;x,y)= T(G-e;x,y) + T(G\backslash e;x,y)$ where $e$ is
neither a loop nor a bridge and $G-e$ and $G \backslash e$ denote
the result of deleting and contracting the edge $e$.

3.- $T(G;x,y)= y T(G-e;x,y)$ when $e$ is a loop

4.- $T(G;x,y)= xT(G/e;x,y)$ when $e$ is a bridge

This are the properties which define the Tutte Polynomial. It
happens that when $x=1, y=1$, the Tutte polynomial of the graph $G$
gives the number of its spanning trees.

\end{document}